\documentclass[review]{elsarticle}

\graphicspath{ {./figures/} }
\usepackage{hyperref}
\usepackage{float}
\usepackage{verbatim} 
\usepackage{apalike}
\usepackage{geometry}
\restylefloat{figure}
\restylefloat{table}
\usepackage{booktabs}
\usepackage{lineno}
\PassOptionsToPackage{hyphens}{url}\usepackage{hyperref}
\usepackage{newtxmath}
\usepackage{url}

\journal{Expert Systems with Applications}

\begin{document}
\begin{frontmatter}

\begin{titlepage}
\begin{center}
\vspace*{1cm}

\textbf{Job recommendations: benchmarking of collaborative filtering methods for classifieds}

\vspace{1.5cm}

Robert Kwieciński$^{a,b}$ (r.kwiecinskipl@gmail.com), Agata Filipowska$^{a,c}$ (agata.filipowska@ue.poznan.pl), Tomasz Górecki$^b$ (tomasz.gorecki@amu.edu.pl), Viacheslav Dubrov (slavadubrov@gmail.com) \\

\hspace{10pt}

\begin{flushleft}
\small  
$^a$ OLX Group, Królowej Jadwigi 43, 61-872 Poznań, Poland \\
$^b$ Adam Mickiewicz University, Wieniawskiego 1, 61-712 Poznań, Poland \\
$^c$ Poznań University of Economics and Business, Al. Niepodległości 10, 61-875 Poznań, Poland

\vspace{1cm}
\textbf{Corresponding Author:} \\
Robert Kwieciński \\
Adam Mickiewicz University, Wieniawskiego 1, 61-712 Poznań, Poland \\
Tel: +48 888 791 384 \\
Email: r.kwiecinskipl@gmail.com

\end{flushleft}        
\end{center}
\end{titlepage}

\title{Job recommendations: benchmarking of collaborative filtering methods for classifieds}

\author[a,b]{Robert Kwieciński \corref{cor1}}
\ead{r.kwiecinskipl@gmail.com}
\author[a,c]{Agata Filipowska}
\ead{agata.filipowska@ue.poznan.pl}
\author[b]{Tomasz Górecki}
\ead{tomasz.gorecki@amu.edu.pl}
\author{Viacheslav Dubrov}
\ead{slavadubrov@gmail.com}

\cortext[cor1]{Corresponding author}
\address[a]{OLX Group, Królowej Jadwigi 43, 61-872 Poznań, Poland}
\address[b]{Adam Mickiewicz University, Wieniawskiego 1, 61-712 Poznań, Poland}
\address[c]{Poznań University of Economics and Business, Al. Niepodległości 10, 61-875 Poznań, Poland}

\begin{abstract}

Classifieds provide many challenges for recommendation methods, due to the limited information regarding users and items. In this paper, we explore recommendation methods for classifieds using the example of OLX Jobs.

The goal of the paper is to benchmark different recommendation methods for jobs classifieds in order to improve advertisements' conversion rate and user satisfaction.

In our research, we implemented methods that are scalable and represent different approaches to recommendation, namely ALS, LightFM, Prod2Vec, RP3beta, and SLIM. 
We performed a laboratory comparison of methods with regard to accuracy, diversity, and scalability (memory and time consumption during training and in prediction). Online A/B tests were also carried out by sending millions of messages with recommendations to evaluate models in a real-world setting.

In addition, we have published the dataset that we created for the needs of our research. To the best of our knowledge, this is the first dataset of this kind. The dataset contains 65,502,201 events performed on OLX Jobs by 3,295,942 users, who interacted with (displayed, replied to, or bookmarked) 185,395 job ads in two weeks of 2020.

We demonstrate that RP3beta, SLIM, and ALS perform significantly better than Prod2Vec and LightFM when tested in a laboratory setting. 
Online A/B tests also demonstrated that sending messages with recommendations generated by the ALS and RP3beta models increases the number of users contacting advertisers. Additionally, RP3beta had a 20\% greater impact on this metric than ALS.
\end{abstract}


\end{frontmatter}

\section{Introduction}\label{intro}


Online classifieds are websites where advertisers post their advertisements concerning the sale or rental of services or products \citep{Davis2021}. The global online classifieds market is forecast to reach US\$32.33 billion in 2024, increasing at a CAGR of 9.40\% between 2020 and 2024 \citep{RM2021}. This paper addresses online advertising using the example of OLX Jobs.

Classifieds make much effort to address the needs of customers, both advertisers and responders to online ads. To match these groups better recommendation methods are gaining in importance. Compared with traditional e-commerce (e.g., Amazon or eBay), classifieds need to address different challenges. Among others, there is the temporal visibility of ads, caused either by limited display duration or by the fact that an advertiser usually needs only one user to complete the transaction.

In our research, we focus on approaches emerging from the field of collaborative filtering. Collaborative filtering models have already proven their accuracy, even though they do not utilize additional information about users or items, which is sometimes hard to obtain or process in practice.
The goal of the paper is to evaluate different recommendation methods for use in the classifieds sector. The high sparsity of our dataset, as well as the significant numbers of ads and users, contribute to the complexity of the research problem.

Our main research contributions are the following: we comprehensively evaluate diverse, scalable collaborative filtering approaches in a laboratory setting, and provide results of online A/B testing carried out with over 1 million users. Additionally, we have published a job interactions dataset. To our best knowledge, this is the largest publicly available dataset on job interactions.

The paper consists of seven sections. Section 2 is a literature review on underlying developments. Section 3 describes the dataset that was developed and made publicly available on Kaggle. The design of the lab experiment, as well as the implementation of methods, are described in section 4. Section 5 focuses on the results of the lab evaluation of our implementations, also addressing the statistical importance of the results. Section 6 presents the results of online implementation performed in the course of A/B tests with users, as well as a discussion of the results. Conclusions are presented in section 7.

\section{Related work}

\subsection{Classified ad sites}

Classified ad sites (classifieds) are online versions of the advertisement sections in newspapers, with ads concerning vehicles, real estate, employment, pets, etc. Classifieds are a special type of e-commerce sites, as they do not support the finalization of transactions between parties -- they are used rather to communicate about the product or service and to set up meetings, while the transaction is carried out offline; this distinguishes them from traditional e-commerce sites such as eBay or Amazon \citep{jones2009online}. 
Our research is focused on OLX, which has a presence in over 30 countries around the world.

One of the ad types published on classifieds is job offers. \cite{Johnson1995} analyzed the format of job ads and how those ads describe the preferences of employers regarding the skills required of candidates. 
In our research, we identified the following characteristics that distinguish job ads on classifieds from, for example, car or real estate ads:

\begin{itemize}
    \itemsep0em 
    \item The requirements set out in an ad indicate the type of user who may respond to the ad -- not all users possess the required competences.
    \item More intensive involvement of the user during the search process, as finding a job is a primary need when compared with, for example, buying a car.
    \item The location of the job has a high impact on the user -- to buy a mobile phone or a car, a user may travel or use a parcel service to deliver the item, but a job usually requires re-location if it is distant from where the user currently lives.
    \item From a technical perspective, a job description has many qualitative aspects, and depending on the position, the differences may be significant even between ads from the same company.
\end{itemize}

On top of these features, in the course of our research we also identified the following challenges for recommendation methods on classifieds that influenced our implementation of the methods:
\begin{itemize}
    \itemsep0em 
    \item Number of users and number of offers: in the case of OLX Jobs we dealt with millions of users and tens of thousands of offers, both constantly changing.
    \item A user does not have to create a profile to interact with ads, meaning that information from a profile cannot be used for recommendation, and recommendations are determined by user behavior during viewing.
    \item An ad usually has a limited number of visitors (interacting with the ad), as an ad usually concerns a unique offer, and after the need is addressed, the ad is disabled.
    \item There is no information on whether a transaction took place; on classifieds, conversion relates to the fact of receiving an answer to a published ad, as any transaction is carried out offline.
\end{itemize}
These challenges create difficulties not only for implementation, but also for testing the accuracy of the proposed methods. We took these challenges into account while preparing the dataset and while implementing the methods described in this paper.

\subsection{Recommendation systems}
\label{section:recommendation_systems}
It has been shown that the problem of \textit{overchoice} \citep{overchoice} can decrease the consumption of products \citep{overchoice_example1, overchoice_example2} and reduce revenue to the site \citep{overchoice_revenue}.
To reduce the number of choices presented, providers can ask their users for additional information about their expectations regarding items. Even if the user is willing to spend time providing this information, the number of possibilities is usually still too large to present all of them to the user. To address this problem, many companies, such as Netflix \citep{netflix} and Amazon \citep{amazon}, have created personalized recommendation systems whose role is to suggest relevant items to users \citep{Aggarwal}.

Recommendations are used in multiple domains, including music \citep{music}, tourism \citep{tourism}, fashion \citep{fashion}, and food \citep{food}. Multiple studies have been made within the domain of e-commerce \citep{recsys-ecommerce, trust_recsys_ecommerce}, classifieds \citep{avito_allegro_recsys} and job recommendations \citep{job_recsys_survey, recsys_challange_2016, recsys_challange_2017}.

Typically we distinguish two approaches for providing recommendations: content-based recommendations and collaborative filtering \citep{Aggarwal}.

In content-based recommendations \citep{content_based_recsys, content_based_recsys2} we utilize information about users (e.g., hobby, education, skills) and items (e.g., price, location, category). These systems usually also make use of the history of interactions (ratings, visits, purchases, replies) of a single user, but do not consider the ratings of all users at the same time.

Collecting and processing information about users and items is not always feasible. In the collaborative filtering approach \citep{recsys_cf_sota1, recsys_cf_sota2} we utilize only information about interactions between users and items. Even in the absence of information on user and item features, these systems are able to provide very accurate recommendations. We have chosen this approach for our problem, because of the lack of user features and the difficulty of outperforming collaborative filtering by utilizing item features in a production environment (we elaborate on this in section \ref{section:discussion}).

One of the greatest shortcomings of collaborative filtering is the impossibility of providing recommendations regarding users and items without any interactions, which is known as the cold-start problem \citep{cold_start_and_feedback_type}. One of the possible solutions to reduce this problem is frequent retraining of the model. In the case of a large number of user interactions, this strategy requires focus on the scalability of the model \citep{scalability}.

Another important factor determining the choice of model is the type of interaction between a user and an item \citep{cold_start_and_feedback_type}. We call the feedback \textit{explicit} when the user explicitly indicates their preference towards the item (e.g., rates it on a scale of 1--5). Another type is \textit{implicit} feedback, when we can only assume the user's preference towards the item based on user behavior (e.g., buying an item, watching a movie). Implicit feedback is less accurate, more abundant, and usually expresses only positive preference \citep{explicit_vs_implicit}.

Laboratory evaluation of recommender systems is usually based on accuracy metrics \citep{metrics_consistency}. Recently, much attention has been focused on other aspects, such as diversity, serendipity, novelty, and coverage \citep{diversity_serendipity_novelty_coverage, diversity}. In the case of classifieds, an item can usually be consumed only by one user, and so we should avoid recommending the same item to a great number of users. Hence, it is important to focus on catalog coverage \citep{coverage}. One of the most challenging tasks in the evaluation of recommender systems is to find the offline metric which is the most correlated with the online goal \citep{indeed-offline-evaluation}.

In this paper, we benchmark collaborative filtering models for a large implicit feedback dataset. We compare the accuracy, diversity and efficiency of selected recommendation methods in a laboratory setting, and then we prove the quality of the methods in online tests with users.

\subsection{Recommendation methods}
When selecting methods for our research, we focused on the following features: proven performance of a method (with a special focus on methods frequently tested in the literature), whether the solution had been tested in production (in real-life scenarios by companies), and scalability (solution able to address millions of users and millions of items). We also wanted to test methods representing different “families” to assess their applicability to classifieds. Finally, ALS, LightFM, Prod2Vec, RP3beta, and SLIM were selected for testing and evaluation. These methods are presented in the following sections.

\subsubsection{Matrix factorization models}

Matrix factorization techniques have been successfully utilized in recommender systems since the Netflix Prize competition \citep{Koren2009}. The main idea is to approximate the sparse user--item interaction matrix as a product of two smaller, dense matrices. We will evaluate two matrix factorization approaches: LightFM and ALS.
\newline

The \textbf{LightFM} \citep{lightfm} model was proposed to overcome the cold-start problem of matrix factorization methods. In our case, without additional information about users and items, it reduces to classical matrix factorization. Despite this, we decided to use this approach, because of its efficient official implementation supporting multiple loss functions: logistic, BPR \citep{bpr}, WARP \citep{warp} and $k$-OS WARP \citep{warp-kos}.

Let $\mathcal{U}$ be a set of users and $\mathcal{I}$ a set of items. For a given user $u \in \mathcal{U}$ and item $i \in \mathcal{I}$ we define $r_{ui} = 1$ if user $u$ interacted with item $i$, and  $r_{ui} = 0$ otherwise. The LightFM model predicts the score $r_{ui}$ as:
$$\hat{r}_{ui} = \pmb{x}_u \cdot \pmb{y}_i + b_u + b_i,$$
where $\pmb{x}_u$ and $\pmb{y}_i$ denote user $u$'s and item $i$'s $k$-dimensional latent representations, and $b_u$ and $b_i$ are user $u$'s and item $i$'s biases.
All of these parameters are learned during the training. 

The logistic loss function should be used when both positive and negative feedback is available, so it is not appropriate for our dataset, which we confirmed experimentally during the hyperparameter tuning. 

The BPR, WARP and $k$-OS WARP loss functions are pairwise learning-to-rank approaches, which are more relevant for our top-$k$ recommendations task. They try to modify model parameters to maximize the difference between the scores of the items $i \in \mathcal{D}_u$ with which user $u$ interacted and of the items $i \notin \mathcal{D}_u$ with which the user did not interact.

Specifically, in the \textbf{BPR loss function} \citep{bpr} we are trying to maximize the expression:
$$
\sum_{(u,i,j) \in \mathcal{S}} \operatorname{ln}\sigma(\hat{r}_{ui}-\hat{r}_{uj}) - \lambda_{\pmb\Theta}||\pmb\Theta||^2,
$$
where
$$
\mathcal{S} = \lbrace (u,i,j)| u \in \mathcal{U}, i \in \mathcal{D}_u  \mbox{ and } j \notin \mathcal{D}_u \rbrace,
$$
$$\sigma(x)=\frac{1}{1+e^{-x}},$$
$\lambda_{\pmb\Theta}$ is a model-specific regularization parameter, $\pmb\Theta$ is a parameter vector, and $||\cdot||$ is the Euclidean norm.

We will present WARP as a special case of $k$-OS WARP.
The $k$-OS WARP loss function is a sum of losses for each user:
$$\sum_{u \in \mathcal{U}} L(\hat{\pmb r}_u, \mathcal{D}_u),$$
where $\hat{\pmb r}_u$ is a vector of estimated scores $\hat{r}_{ui}$ for each item $i \in \mathcal{I}$.

For a given user $u$ we order the interacted items $\mathcal{D}_u$ in decreasing order of the estimated scores. Namely, we order the items as $u_{i_1}, u_{i_2}, \ldots, u_{i_{|\mathcal{D}_u|}}$, where the indices $i_k$ are chosen in such a way that:
$$
\hat{r}_{ui_1} \geq \hat{r}_{ui_2} \geq \ldots \geq \hat{r}_{ui_{|\mathcal{D}_u|}}.
$$
Then:
$$
L(\hat{\pmb r}_u, \mathcal{D}_u) = \frac{1}{Z}\sum_{s=1}^{|\mathcal{D}_u|} P(s) \Phi(\operatorname{rank}_{i_s}(\hat{\pmb r}_u)),
$$
where $\operatorname{rank}_{i}(\hat{\pmb r}_u) = \sum_{j \in \mathcal{I} \setminus \mathcal{D}_u} \vmathbb{1}(1 + \hat{r}_{uj} \geq \hat{r}_{ui})$, $\vmathbb{1}$ is an indicator function, $\Phi(n) = \sum_{m=1}^n \frac{1}{m}$, and $Z=\sum_{s=1}^{\mathcal{|D}_u|} P(s)$ normalizes the weights induced by $P$.

In practice, we sample a positive item $i$ with respect to the weighting function $P$ and randomly pick unseen items $j$ until, after $N$ samplings, $1 + \hat{r}_{uj} \geq \hat{r}_{ui}$. Then we perform a gradient step to minimize $\Phi(\lfloor\frac{|\mathcal{I} \setminus \mathcal{D}_u|}{N}\rfloor)(1 - \hat{r}_{ui} + \hat{r}_{uj})$.

Finally, the $k$-OS WARP loss is achieved for $P(k)=1$ and $P(m)=0$ if $m \neq k$. The WARP loss is achieved when we set $P(m)=1$ for $m \in \mathbb{N}$. In this case, to simplify the computations, we can skip ordering positive items from $\mathcal{D}_u$.

In our research, all loss functions (logistic, BPR, WARP and $k$-OS WARP) were treated as hyperparameters of the LightFM model and optimized during training.
\\

The \textbf{ALS} model (Alternating Least Squares), also known as WRMF -- Weighted Regularized Matrix Factorization \citep{Yifan2008}, is an example of a matrix factorization method designed for implicit feedback datasets.
We chose this method because of its proven performance, scalability and popularity in research and industry. Additionally, the ALS model was already implemented and used in OLX.

We define the \textbf{confidence} of an observation as $c_{ui}=1+\alpha r_{ui}$, where $\alpha$ is a hyperparameter.
We also define the \textbf{preference} as
$p_{ui}=\vmathbb{1}(r_{ui} > 0)$.
Then, in ALS we optimize the following expression:
$$\min_{\pmb{x}_*,\pmb{y}_*}\sum_{u,i} c_{ui}(p_{ui}-\pmb{x}^\top_u \pmb{y}_i)^2+\lambda(\sum_u ||\pmb{x}_u||^2+\sum_i ||\pmb{y}_i||^2),$$
where $\pmb{x}_u$ is user $u$'s $f$-dimensional embedding, $\pmb{y}_i$ is item $i$'s $f$-dimensional embedding, and $\lambda$ is a regularization parameter. The name of the model refers to the method for which this function is optimized. The Alternating Least Squares procedure assumes that either user or item latent factors are fixed, and analytically determines the others. The procedure is repeated a few times in a very efficient way \citep{Yifan2008,Gabor2011}.

\subsubsection{Neighborhood-based models}

The \textbf{SLIM} model (Sparse Linear Methods for Top-N Recommender Systems; \cite{Ning2011}) is based on item-based nearest neighbors regression \citep{Aggarwal}. We chose this method because of its advantages over item-based $k$-nearest-neighbors and matrix factorization models \citep{Ning2011} and good performance against non-neural and neural approaches \citep{Dacrema2021}.

In this model we approximate the user--item matrix $\pmb{R}$ by learning the item--item sparse similarity matrix $\pmb{W}$. Explicitly, $\pmb{W}$ is learned by minimizing the expression
$$\frac{1}{2}||\pmb{R}-\pmb{RW}||^2_F+\frac{\beta}{2}||\pmb{W}||^2_F+\lambda||\pmb{W}||_1,$$
where $w_{ij} \geq 0$ and $w_{ii}=0$ for all $i$, $j$, $1 \leq i,j \leq |\mathcal{I}|$,\\
$||\cdot||_F$ is the matrix Frobenius norm:
$$||\pmb{W}||_F=\sqrt{\sum_{i=1}^{|\mathcal{I}|}\sum_{j=1}^{|\mathcal{I}|}w_{ij}^2},$$\\
$||\cdot||_1$ is the entry-wise $\ell_1$-norm: $$||\pmb{W}||_1=\sum_{i=1}^{|\mathcal{I}|}\sum_{j=1}^{|\mathcal{I}|}|w_{ij}|$$
and $\beta$ and $\lambda$ are regularization parameters.
\newline

We can see that each column of $\pmb{W}$ can be learned independently, which improves the scalability of this model \citep{Ning2011}.

\subsubsection{Graph-based approaches}

\textbf{RP3beta} \citep{Christoffel} is a graph-based collaborative filtering approach. Similarly to \cite{Dacrema2021}, we will present this model as an extension of the P3alpha model \citep{p3alpha}. 
We consider a bipartite and undirected graph $G = (\mathcal{V}, \mathcal{E})$, where each vertex $v \in \mathcal{V}$ represents either a user or item, and each edge $\lbrace v_u, v_i \rbrace \in \mathcal{E}$ represents the interaction between a given user $u$ and a given item $i$.
Let $\pmb{A}$ be the adjacency matrix and $\pmb{D}$ the degree matrix of the graph $G$.
Then the scores of P3alpha are stored in the matrix 
$$((\pmb{D}^{-1}\pmb{A})^{\circ \alpha})^3,$$
where $\circ$ denotes the element-wise power of a matrix (Hadamard power).
More precisely, the score $\hat{r}_{ui}$ is stored on the intersection of the row representing user $u$ and the column representing item $i$. We can see that this score is a sum of scores assigned to paths of length 3 connecting user $u$ and item $i$ in the graph $G$.

In RP3beta, the score is computed with P3alpha and divided by the popularity of the items raised to the power of $\beta$, to mitigate the popularity bias.

We decided to choose this model because of its simplicity, scalability and proven performance \citep{Dacrema2021, reenvisioningNCF}, especially in view of the high sparsity of our dataset, which significantly decreases memory utilization when training the model. This enables us to directly compute and store in memory the item--item similarity matrix, instead of the random walks approximation proposed in the original paper.

\subsubsection{Prod2Vec}

The \textbf{Prod2Vec} \citep{Grbovic} (or Item2vec \citep{item2vec}) model is based on the Word2Vec model \citep{Mikolov2013} widely used in Natural Language Processing. The model focuses on learning vector representations of items using a sequence of items as a “sentence'' and items within that sequence as “words''. This method was chosen because it is scalable and very different from the compared methods. Additionally, the current state-of-the-art hybrid deep neural network recommender system used in other categories at OLX is based on this approach.

Prod2Vec uses the skip-gram model by maximizing the objective function over the set $\mathcal{S}$ of item sequences, defined as follows:
$$L = \sum_{s \in \mathcal{S}} \sum_{p_i \in s} \sum_{\substack{-c \leq j \leq c \\ j \neq 0}} \operatorname{log}\mathbb{P}(p_{i+j}|p_i),$$

\noindent where $c$ is the length of the context for item sequences, and items from the same sequence are ordered arbitrarily. The probability $\mathbb{P}(p_{i+j}|p_i)$ of observing a neighboring item $p_{i+j}$ given the current item $p_i$ is defined using the soft-max function:
$$\mathbb{P}(p_{i+j}|p_{i}) = \frac{\operatorname{exp}(\pmb{v}_{p_i}^\top \pmb{v}_{p_{i+j}}^{'})}{\sum_{p=1}^P\operatorname{exp}(\pmb{v}_{p_i}^\top\pmb{v}_p^{'})},$$
\noindent where $\pmb{v}_{p}$ and $\pmb{v}_{p}^{'}$ are the input and output vector representations of item $p$, and $P$ is the number of unique items in the vocabulary. Prod2Vec models the context of an item sequence, where items with similar contexts (i.e., with similar neighboring interactions) have similar vector representations.

To generate recommendations, we represent the user as an average of input vector representations of interacted items. Then we search for items whose input representations have the highest cosine similarity with the user representation ($k$-nearest neighbors in the latent space).

\subsection{Research gap}

The goal of the paper is to benchmark different methods in the jobs domain on classifieds, which is a targeted research gap. To the best of our knowledge, the recommendation methods addressed have never been either tested or benchmarked on a large scale and while working online in the jobs domain. In this work, we compare these methods not only in a laboratory setting using the dataset published by us, but also in online A/B tests carried out in a production environment.


\section{Dataset}

This research provides a commercial dataset that contains 65,502,201 events performed on OLX Jobs by 3,295,942 users who interacted with 185,395 job ads in two weeks of 2020. We have published the dataset on Kaggle.\footnote{\url{https://www.kaggle.com/olxdatascience/olx-jobs-interactions}, accessed: 2022-08-06}
  
\begin{table}[ht]
\footnotesize
\centering
\caption{Example rows from the dataset.}
\label{dataset}
\begin{tabular}{llll}
\toprule
User   & Item   & Event & Timestamp  \\
\midrule
1745587  & 168661  & click & 1582216025 \\
843008 & 62838 & click & 1582485868 \\
14285 & 30469   & bookmark & 1582247367 \\
1142944 & 80122 & click & 1581805847 \\
2835659 & 23728  & chat\_click & 1582397836 \\
\bottomrule
\end{tabular}
\\ Source: own study
\end{table}

Each row of the dataset represents the event performed by a given user regarding a given ad at a given time (Table \ref{dataset}). The following events involving user behavior were included:
\begin{itemize}
    \itemsep0em 
    \item click: the user visited the item detail page;
    \item bookmark: the user added the item to bookmarks;
    \item chat\_click: the user opened the chat to contact the item’s owner;
    \item contact\_phone\_click\_1: the user revealed the phone number attached to the item;
    \item contact\_phone\_click\_2: the user clicked to make a phone call to the item’s owner;
    \item contact\_phone\_click\_3: the user clicked to send an SMS message to the item’s owner;
    \item contact\_partner\_click: the user clicked to access the item’s owner's external page;
    \item contact\_chat: the user sent a message to the item’s owner.
\end{itemize}

\begin{figure*}[ht]
    \centering
  \resizebox{.9\textwidth}{!}{
    \includegraphics{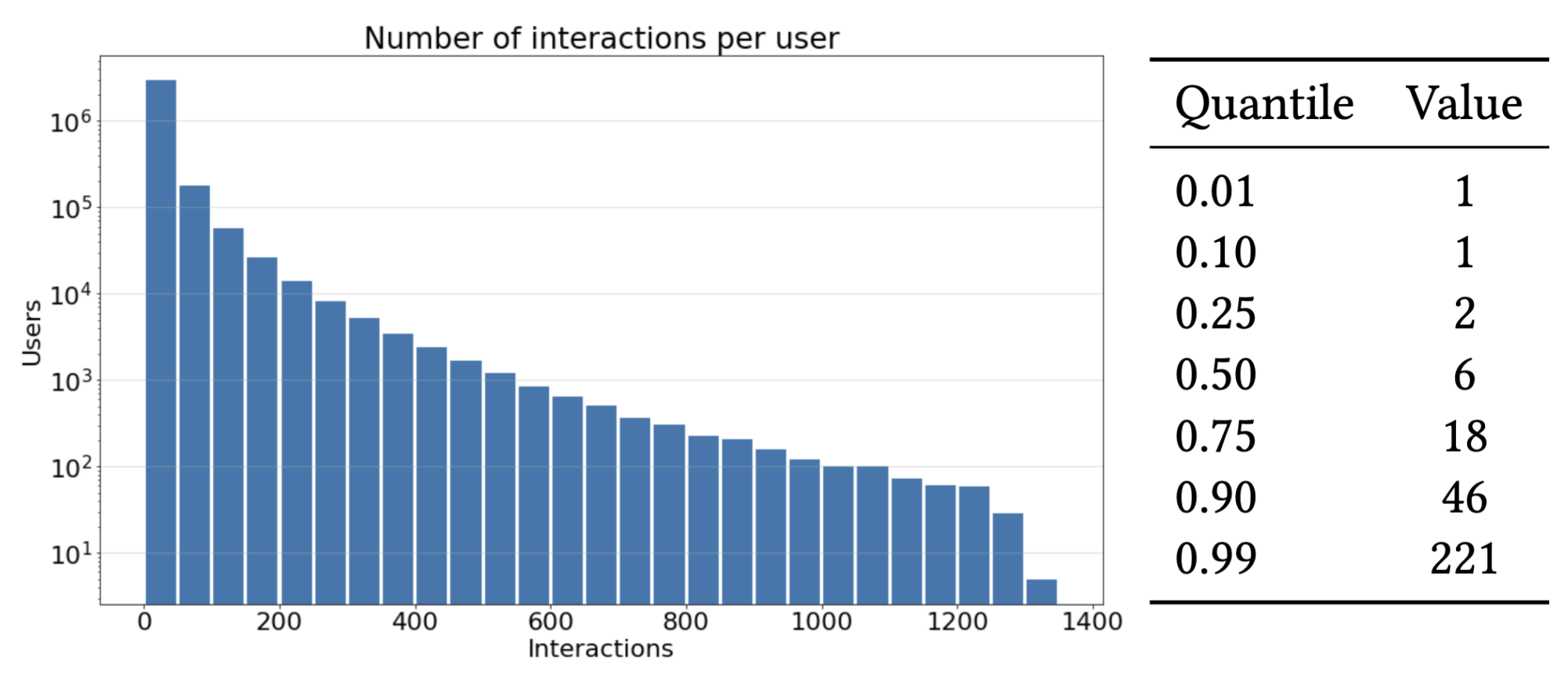}}
    \caption{Number of interactions per user. Source: own study}
    \label{int_user}
  \end{figure*}

\begin{figure*}[ht]
    \centering
    \resizebox{.9\textwidth}{!}{
    \includegraphics{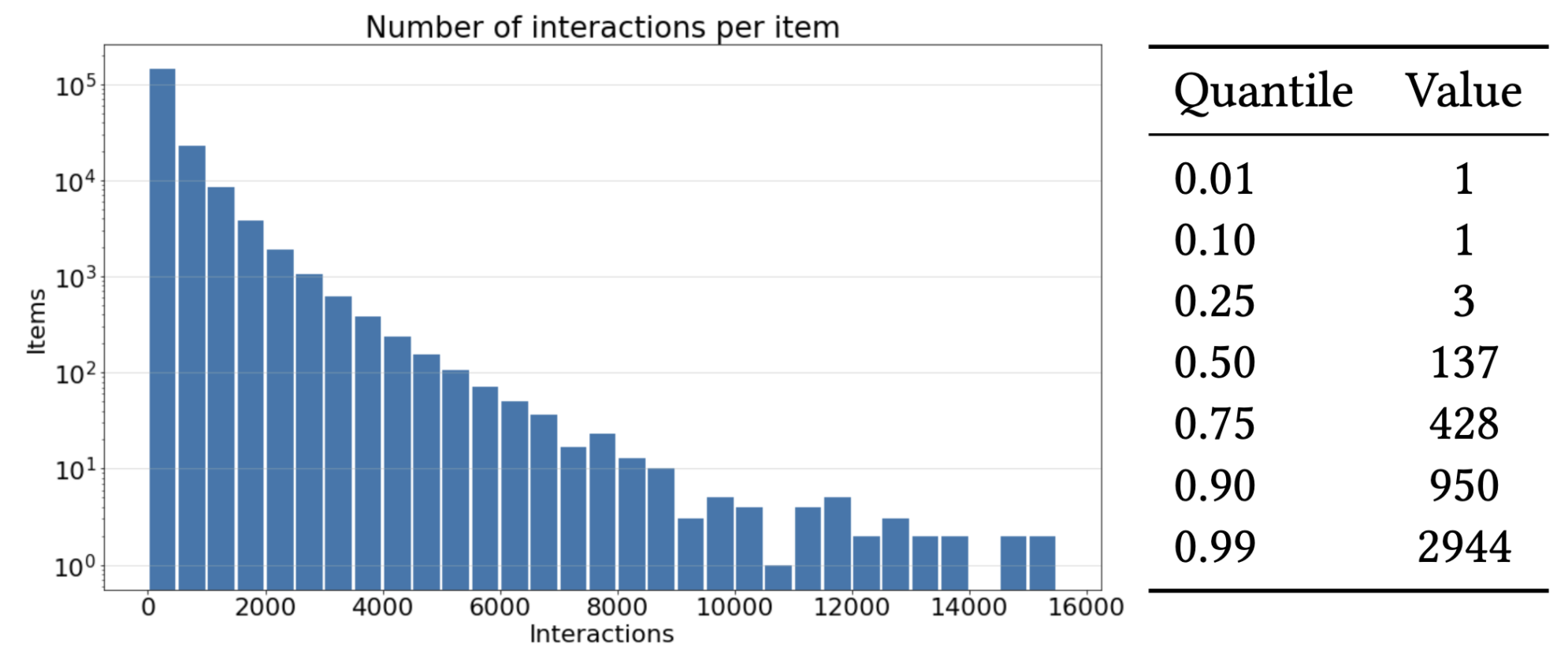}}
    \caption{Number of interactions per item. Source: own study}
    \label{int_item}
\end{figure*}

The data sparsity (\% of missing entries in the user--item matrix) equals 99.9893\%, which is very low, but typical for classifieds (for comparison, the data sparsity on the MovieLens 20M Dataset is 99.47\%, making it almost 50 times denser). The average number of interactions per user is 19.874 (with a standard deviation of 47.762 and a maximum of 1310). The average number of interactions per item is 353.312 (with a standard deviation of 624.773 and a maximum of 15,480).
The statistics for the dataset are provided in Figures \ref{int_user} and \ref{int_item}, as well as in Table \ref{events}. Note that we used a logarithmic scale, because the distributions are strongly right-skewed.

\begin{table}[ht]
\footnotesize
\centering
\caption{Distribution of events.}
\label{events}
\begin{tabular}{lc}
\toprule
Action                   & Frequency (\%)\\
\midrule
click                    & 89.794  \\
contact\_phone\_click\_1   & 2.628   \\
bookmark                 & 2.511   \\
chat\_click              & 2.136   \\
contact\_chat            & 1.448   \\
contact\_partner\_click  & 0.701   \\
contact\_phone\_click\_2 & 0.679   \\
contact\_phone\_click\_3 & 0.103  \\
\bottomrule
\end{tabular}
\\ Source: own study
\end{table}

It should be noted that maintaining the confidentiality of ads and users was a priority when preparing this dataset. The measures taken to protect privacy included the following:

\begin{itemize}
    \itemsep0em 
    \item original user and item identifiers were replaced by unique random integers;
    \item some undisclosed constant integer was added to all timestamps;
    \item some fraction of interactions were filtered out;
    \item some additional artificial interactions were added.
\end{itemize}


\section{Experimental setup}

\subsection{Implementation}

We implemented the above-mentioned methods based on previous developments. Details on how our implementations differ from the previously tested approaches are provided in Table \ref{sota_methods}. The implementation of the methods for the purpose of reproducibility of the results is available on Github.\footnote{\url{https://github.com/rob-kwiec/olx-jobs-recommendations}, accessed: 2022-08-06}

\begin{table*}[ht]
\footnotesize
    \centering
  \caption{Methods re-implemented for the needs of the research.}
  \label{sota_methods}
  \begin{tabular}{p{1cm}p{1.65cm}p{3.7cm}p{4.1cm}}
    
Method   & Family \newline of methods    & Source  & Difference from the original method      \\ \hline

LightFM     
& Matrix factorization 
& Implementation based on: \citep{LightFM-github}. Supporting paper: \citep{lightfm}.
& None    \\           \hline 

ALS      
& Matrix factorization 
& Implementation based on: \citep{ALS}. Supporting papers: \citep{Gabor2011, Yifan2008}.             
& None                           \\ \hline

SLIM     
& Neighborhood-based  
& Sources that inspired our implementation: \citep{SLIM2, SLIM1}. Supporting paper: \citep{Ning2011}.
& None      \\           \hline 

RP3beta  
& Graph-based          
& Source that inspired our implementation: \citep{RP3beta}. Supporting papers: \citep{Paudel2016, Dacrema2021}.
& Performed direct computations on sparse matrices instead of random walks approximation. \\ \hline

Prod2Vec 
& Word2Vec     
& Implementation based on: \citep{Prod2Vec}. Supporting papers: \citep{Grbovic, item2vec}.
& Sequences of interactions, ordered by a timestamp. Representing user as an average of interacted items' representations. Experimented also with CBOW (Continuous Bag of Words).                                                          \\ \hline

\end{tabular}
\\ Source: own study
\end{table*}

\subsection{Laboratory research and online testing goals}
Our ultimate business goal was to increase the number of users applying for jobs through OLX Jobs. To achieve that, we try to solve a ranking problem; more precisely, we try to recommend 10 items with the highest chance of receiving a given user's interaction. The problem of finding an offline metric that is the most correlated with the apply-rate in the jobs domain was considered by \cite{indeed-offline-evaluation}.
Since our goal is very similar, we decided to follow their results by optimizing for precision@10 (precision for the first 10 recommended items \citep{metrics_consistency}). However, we also report and discuss several other accuracy, diversity and efficiency metrics.

We are focused on users who are actively looking for a job by interacting with job ads. For that reason we do not consider users without any interactions (cold-start users).
\subsection{Data split}

\begin{figure*}[ht]
    \centering
    \resizebox{.9\textwidth}{!}{
    \includegraphics{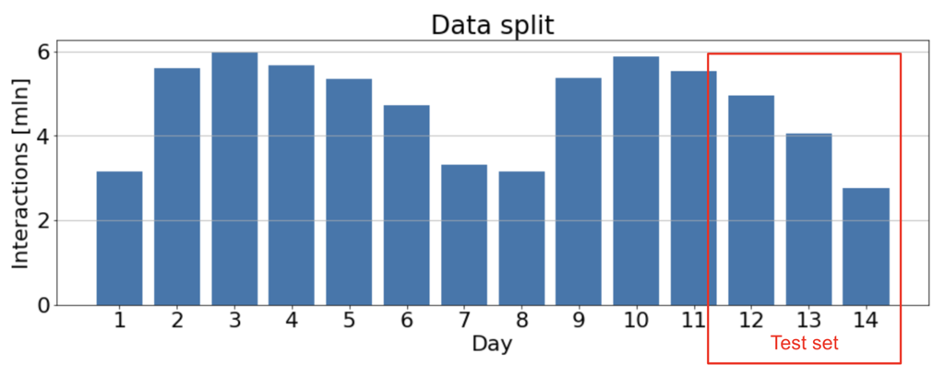}}
    \caption{Dataset preparation (split into training \& validation and test depending on the number of days). Source: own study}
    \label{split}
  \end{figure*}

The dataset described in section 3 was split into training \& validation and test datasets as required by the research methodology. The test set includes 20\% of the newest interactions. This means that out of 14 days, approximately 3 days were included in the test set (see Figure \ref{split}).

Then, in both datasets, we limited the number of interactions to unique interactions between a user and an item. If a user interacted more than once with an item, only the first interaction was counted and the timestamp of that interaction was associated with the item. We also did not distinguish between types of interactions. 
In addition, two more modifications were applied: only users who appeared in the training \& validation dataset were included in the test set (because we could not provide personalized recommendations to other users without additional knowledge about them), and we filtered out all user--item pairs that were present in the training set (in order to avoid recommending items that had already been seen). As a result, we ended up with about 38M rows in the training set and about 6M rows in the test set.

\subsection{Model tuning}

For the sake of cost and time, when performing the optimization of parameters we made a restriction to 20\% of users and 20\% of items of the training \& validation set. Then, we split the restricted dataset into training and validation datasets with the rules described in the section above. For each model, there were 100 iterations chosen by Bayesian optimization using Gaussian processes implemented in ScikitLearn.\footnote{\url{https://scikit-optimize.github.io/stable/modules/generated/skopt.gp_minimize.html}, accessed: 2022-08-06} The configuration file defines spaces for hyperparameters and is available in our code repository.\footnote{\url{https://github.com/rob-kwiec/olx-jobs-recommendations/blob/main/src/tuning/config.py}, accessed: 2022-08-06}  For example, for RP3beta we restricted ourselves to searching for $\alpha$ and $\beta$ in the interval [0,2].

\begin{figure*}[ht]
    \centering
    \resizebox{.9\textwidth}{!}{
    \includegraphics{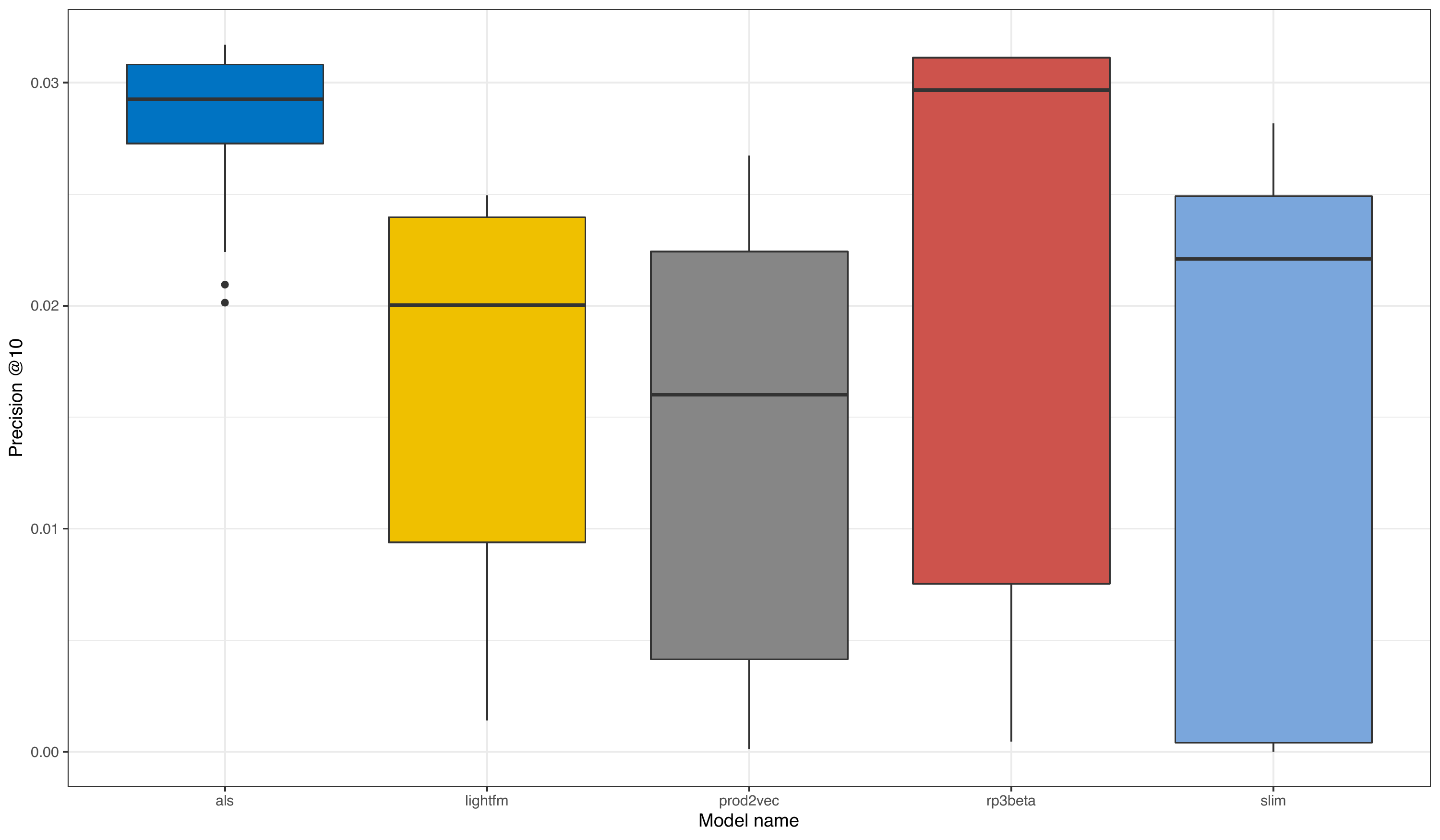}}
    \caption{Precision for each model depending on parameters. Source: own study}
    \label{pic:precision}
  \end{figure*}

Recommendations in the optimization process were prepared for 30,000 users, and as mentioned above, the optimization score used was precision@10. 
Figure \ref{pic:precision} shows the precision score depending on the parameters. It is visible that within the process ALS was not highly optimized, but for other methods, this optimization was crucial to improving precision. Hyperparameters for all models are presented in Table \ref{hyperparams}.

\begin{table*}[ht]
\footnotesize
    \centering
\caption{Model hyperparameters}
\begin{tabular}{lp{9.9cm}}
\toprule
\textbf{Model} & \textbf{Model hyperparameters} \\
\midrule
\textbf{als} & \{'factors': 357, 'regularization': 0.001, 'iterations': 20, 'event\_weights\_multiplier': 63\} \\
\textbf{lightfm} & \{'no\_components': 512, 'k': 3, 'n': 20, 'learning\_schedule': 'adadelta', 'loss': 'warp', 'max\_sampled': 61, 'epochs': 11\} \\
\textbf{prod2vec} & \{'vector\_size': 168, 'alpha': 0.028728, 'window': 20, 'min\_count': 16, 'sample': 0.002690026, 'min\_alpha': 0.0, 'sg': 1, 'hs': 1, 'negative': 200, 'ns\_exponent': -0.16447846705441527, 'cbow\_mean': 0, 'epochs': 22\} \\
\textbf{rp3beta} & \{'alpha': 0.61447198, 'beta': 0.1443548\} \\
\textbf{slim} & \{'alpha': 0.00181289, 'l1\_ratio': 0.0, 'iterations': 3\} \\
\bottomrule
\end{tabular}\label{hyperparams}
\\ Source: own study
\end{table*}

\section{Laboratory evaluation}

In this section we describe the laboratory evaluation of the discussed methods on the developed dataset.
We used the best-performing hyperparameters found in the previous section. Each model was trained on the full training \& validation dataset, and recommendations were produced for all 619,389 users from the test set. To better present the values of each offline metric, we compared the described methods with simple, non-personalized baselines: the most popular product approach and random recommendations.

\subsection{Accuracy}

\subsubsection{Evaluation of accuracy}

All metrics presented in this section consider the first 10 recommendations, e.g., precision@10, recall@10. Higher metric values indicate a higher quality of recommendations. The values should not be directly compared with the results achieved on other datasets, because metrics heavily depend on the distribution of the dataset (for example high sparsity) and the train/test splitting strategy. Table \ref{accuracy} presents the outcomes of the evaluation.
Most of the observed differences are statistically significant, due to the large number of evaluated users. We elaborate more on the statistical significance of the results in section \ref{section:statistics_accuracy}.

\begin{table*}[ht]
\footnotesize
\centering
\caption{Accuracy: evaluation results. All presented metrics were described by \cite{metrics_consistency}. For the mAP metric we used the variant with $x=\min(k,r)$.}
\label{accuracy}
\begin{tabular}{llllllll}
\toprule
Model & RP3beta & SLIM   & ALS    & Prod2Vec & LightFM &  Most popular & Random  \\
\midrule
precision    &\textbf{0.0484}  & 0.0472 & 0.0434 & 0.0368   & 0.0359 & 0.0012       & 0.00006 \\
recall & \textbf{0.0783}  & 0.0736 & 0.0657 & 0.0580 & 0.0564 & 0.0012       & 0.00005 \\
ndcg & \textbf{0.0759}  & 0.0721 & 0.0657 & 0.0567 & 0.0545 & 0.0016       & 0.00007 \\
mAP & \textbf{0.0393}  & 0.0365 & 0.0329 & 0.0282 & 0.0264 & 0.0006       & 0.00002 \\
MRR & \textbf{0.1365}  & 0.1314 & 0.1230 & 0.1065 & 0.1034 & 0.0038       & 0.00019 \\
LAUC & \textbf{0.5391}  & 0.5368 & 0.5328 & 0.5289  & 0.5281 & 0.5006       & 0.49999 \\
HR & \textbf{0.3131}  & 0.3066 & 0.2878 & 0.2537 & 0.2547 & 0.0112       & 0.00059\\
\bottomrule
\end{tabular}
\\ Source: own study
\end{table*}

We can see that all of our approaches greatly outperform both random and most popular item recommendations. The low performance of the most popular approach results from the nature of classifieds, especially the jobs domain, where users are mostly interested in a specific location and a category. In such a scenario it is reasonable to build more sophisticated, personalized recommendation systems.

It is demonstrated that RP3beta outperforms other approaches in terms of accuracy metrics. We believe that this is related to the high sparsity of our dataset. RP3Beta calculates the recommendations in a deterministic way (by leveraging paths of length 3 on a user--item bipartite graph). All other approaches utilize machine learning techniques for finding user representations (ALS, LightFM), item representations (ALS, LightFM, Prod2Vec) or direct item--item similarities (SLIM), which might be challenging in case of users or items with a low number of interactions.

\subsubsection{Statistical comparison for precision@10}
\label{section:statistics_accuracy}
To identify differences between the methods, we present a detailed statistical comparison for our main metric, precision@10. To begin with, we test the null hypothesis that all methods perform the same and the observed differences are merely random (omnibus test). The Friedman test \citep{Friedman37,Friedman40} with the \cite{Iman80} extension is probably the most popular omnibus test, and it is usually a good choice when comparing more than five different algorithms \citep{Garcia08,Garcia2010}. Let $R_{ij}$ be the rank of the $j$th of $K$ methods on the $i$th of $N$ samples and  $$R_j=\frac{1}{N}\sum_{i=1}^{N}R_{ij}.$$ The test compares the mean ranks of methods and is based on the statistic
\begin{align*}
F_F = \frac{(N - 1)\chi^2_F}{N(K - 1) - \chi^2_F},
\end{align*}
where
\begin{align*}
\chi^2_F = \frac{12N}{K(K + 1)}\sum_{i = 1}^KR_i^2 - 3N(K + 1)
\end{align*}
is the Friedman statistic, which has the $F$ distribution with $K - 1$ and $(K - 1)(N - 1)$ degrees of freedom. The $p$-value from this test is equal to 0. The obtained $p$-value indicates that we can safely reject the null hypothesis that all of the algorithms perform the same. We can therefore proceed with the post hoc tests in order to detect significant pairwise differences among all of the methods. \cite{Demsar06} proposes the use of Nemenyi's test \citep{Nemenyi63}, which compares all algorithms pairwise. For a significance level $\alpha$ the test determines the critical difference (CD). If the difference between the average rankings of two algorithms is greater than $$\text{CD} = q_\alpha\sqrt{\frac{K(K + 1)}{6N}}$$ the null hypothesis that the algorithms have the same performance is rejected (the $q_\alpha$ are based on the Studentized range statistic divided by $\sqrt{2}$). \cite{Demsar06} proposes a plot to visually check the differences, the CD plot. In the plot, those algorithms that are not joined by a line can be regarded as different.

\begin{figure}[ht]
\centering
\resizebox{0.7\textwidth}{!}{
\includegraphics{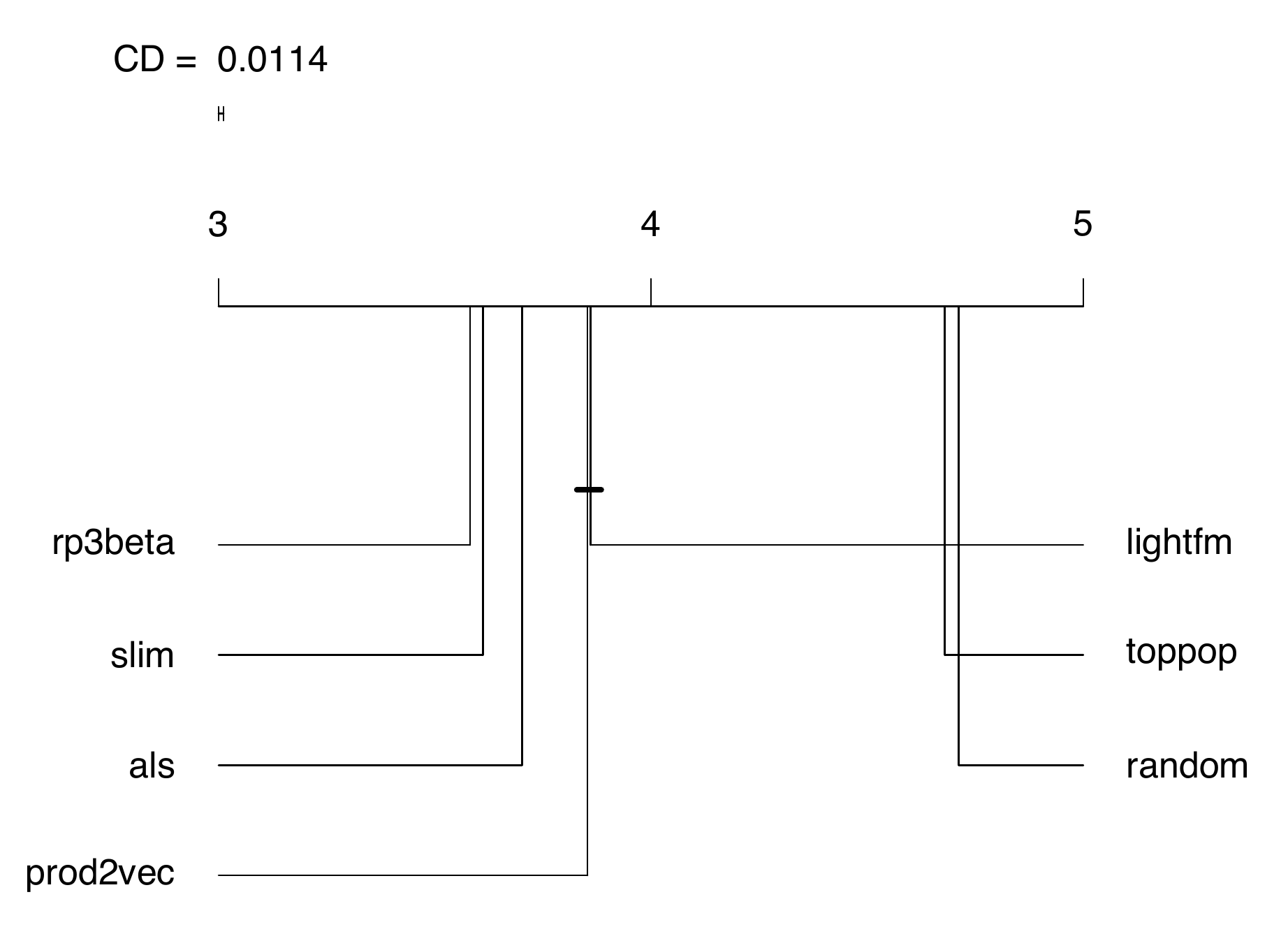}}
\caption{Critical difference plot (for precision). Source: own study}
\label{fig:cdplot}
\end{figure}

In our case, with a significance of $\alpha = 0.05$, any two algorithms with a difference in the mean rank above 0.0114 will be regarded as non-equal (Figure~\ref{fig:cdplot}). 

\subsubsection{Accuracy for users with different numbers of interactions}

\begin{figure*}[ht]
\centering
  \resizebox{.9\textwidth}{!}{
\includegraphics{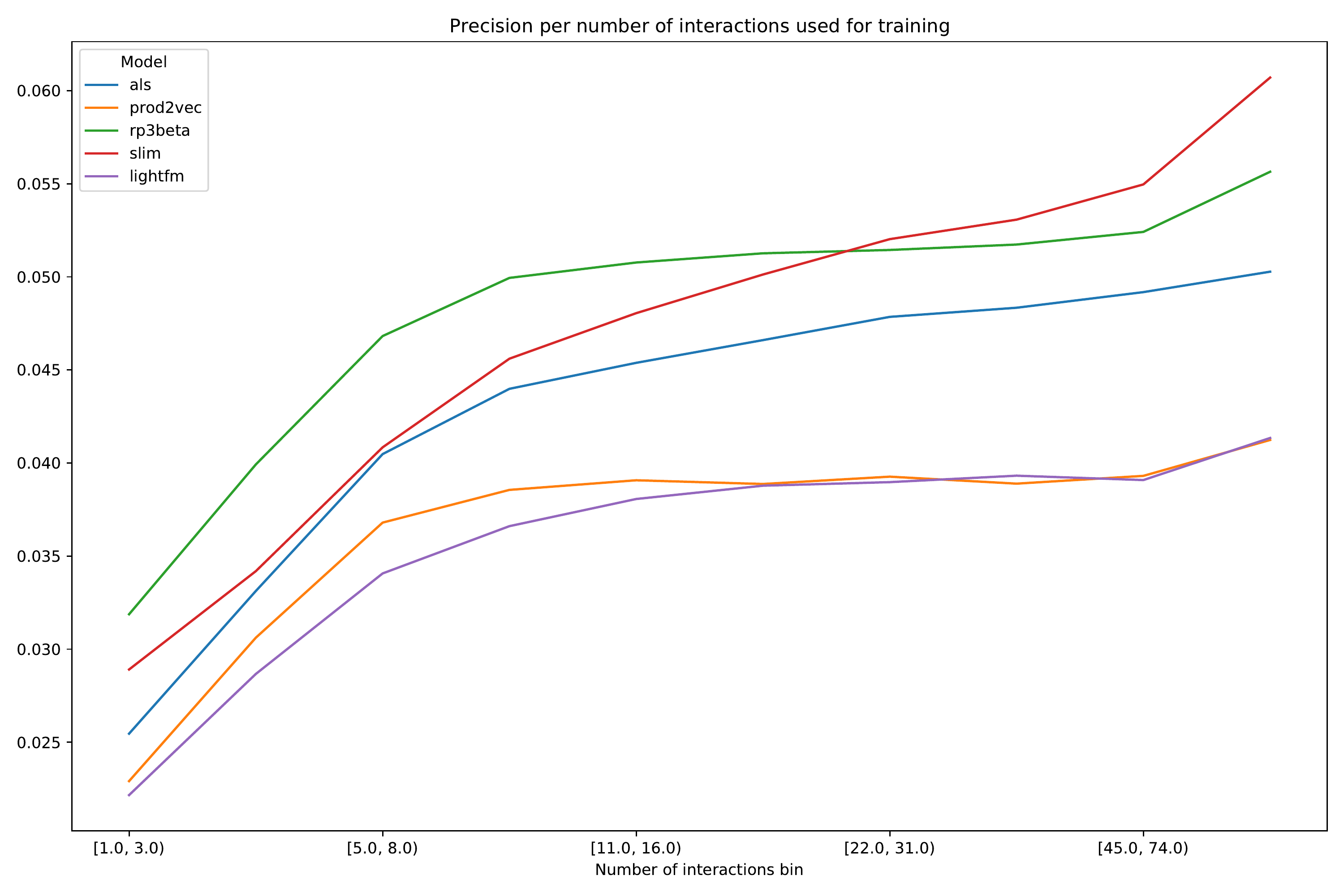}}
\caption{Precision@10 for each model depending on the number of items with which the user interacted. Best viewed in color. Source: own study}
\label{precision_per_nb_inter}
\end{figure*}

The only information we have about our users is their interactions. In Figure \ref{precision_per_nb_inter} we checked how the number of items with which the user interacted influences our main metric, precision@10. We divided our users into 10 groups of similar size depending on the number of interactions.
We can see that the order of models sorted by precision@10 does not depend on the group, except that SLIM performed better than RP3Beta for users with at least 22 interactions. The difference is statistically significant for each group of users (discussed in section \ref{section:statistics_accuracy_bins}). The RP3Beta model can be seen as a special case of the item-based collaborative filtering approach, which uses a deterministic similarity measure between items \citep{Dacrema2019}. On the other hand, SLIM uses machine learning methods for learning similarities. We suppose that such learned similarity scores are less biased, but have higher variance, which leads to better performance for users with many interactions because averaging is done over a greater number of scores.

\subsubsection{Statistical evaluation of accuracy for users with a different number of interactions}
\label{section:statistics_accuracy_bins}

In this subsection we discuss the difference between the RP3Beta and SLIM models depending on the number of user interactions, as observed in Figure \ref{precision_per_nb_inter}.

To statistically compare two methods over multiple datasets, \cite{Demsar06} recommends the Wilcoxon signed-ranks test \citep{Wilcoxon1945}. The Wilcoxon signed-ranks test is a non-parametric alternative to the paired $t$-test, which ranks the differences in performances of two algorithms for each dataset, ignoring the signs, and compares the ranks for the positive and the negative differences.

In Table \ref{statistics-bins} we can observe that, for each group of users, the difference between the RP3Beta and SLIM models is statistically significant.

\begin{table}[ht]
\footnotesize
    \centering
\caption{Results of Wilcoxon signed-ranks test between RP3Beta and SLIM models depending on the number of items with which the user interacted.}
\label{statistics-bins}
\begin{tabular}{ll}
\toprule
\textbf{bin} & \textbf{p-value} \\
\midrule
{[}1.0, 3.0) & 0 \\
{[}3.0, 5.0) & 0 \\
{[}5.0, 8.0) & 0 \\
{[}8.0, 11.0) & 0 \\
{[}11.0, 16.0) & 0 \\
{[}16.0, 22.0) & 1.25e-6 \\
{[}22.0, 31.0) & 9.27e-7 \\
{[}31.0, 45.0) & 0 \\
{[}45.0, 74.0) & 0 \\
{[}74.0, 852.0) & 0 \\
\bottomrule
\end{tabular}
\\ Source: own study
\end{table}

\subsection{Diversity}
\begin{table*}[ht]
\footnotesize
    \centering
\caption{Diversity: evaluation results.}
\label{diversity}
\begin{tabular}{llllllll}
\toprule
Model    & RP3beta & SLIM   & ALS    & Prod2Vec & LightFM & Most popular & Random  \\
\midrule
test coverage & 0.5725  & 0.5171 & 0.3038 & \textbf{0.7400}   & 0.7031 & 0.0002       & 0.9778  \\
Shannon & 9.5271  & 9.6728 & 9.6270 & \textbf{10.4031} & 10.1385 & 2.3296       & 11.7267 \\
Gini & 0.9083  & 0.9029 & 0.9120 & \textbf{0.7956} & 0.8397 & 0.9999       & 0.1159 \\
\bottomrule
\end{tabular}
\\ Source: own study
\end{table*}

In the classifieds domain, especially the jobs domain, usually only one user is needed to complete a transaction with respect to a given ad. In this case the diversity of recommendations is even more important than in other domains (like music, movies, or even other e-commerce excluding classifieds).
To assess this aspect, we use three measures: test coverage, Shannon Entropy \citep{Shani2011}, and the Gini Index \citep{Shani2011}. Test coverage refers to the fraction of test items that were recommended to at least one user. In the case of Shannon Entropy and the Gini Index we ignore items outside the test set. Greater values of test coverage and Shannon Entropy indicate higher diversity, whereas greater values of the Gini Index indicate lower diversity.
In Table \ref{diversity} we can observe that Prod2Vec and LightFM provide the greatest diversity, after excluding random recommendations (which by definition should have the greatest diversity). The diversities of the most accurate models, RP3Beta, SLIM and ALS, are similar, except for the case of the lower test coverage of ALS.

\subsection{Overlap of methods}

In our research, we focus on methods that are worth testing with real users. For that reason, we studied the overlap between recommendations offered by the implemented methods with respect to user--item pairs (Overlap Coefficient \citep{Vijaymeena2016}). We expect more significant differences in recommendations when comparing methods with a low overlap.\\

The results of this experiment are reported in Table \ref{overlap}.
RP3beta and SLIM offered similar recommendations. This is because they both use a sparse item--item similarity matrix, although calculated differently. The group of comparisons with ALS has medium similarities with other methods. The most different from the overlap perspective are also the most diverse models, namely Prod2Vec and LightFM. The low overlap of ALS and LightFM emphasizes the importance of the choice of the loss function in matrix factorization approaches.

\begin{table}[ht]
\footnotesize
    \centering
\caption{Overlap of models.}
\label{overlap}
\begin{tabular}{llllll}
\toprule
 Model         & RP3beta & SLIM  & ALS & Prod2Vec & LightFM \\
\midrule
RP3beta  & 100\%   & \textbf{73\%}& 53\%  & 37\% & 38\%     \\
SLIM     & \textbf{73\%}   & 100\% & 50\% & 35\% & 35\%     \\
ALS      & 53\%    & 50\%  & 100\% & 38\% & 37\%     \\
Prod2Vec & 37\%    & 35\%  & 38\%  & 100\% & 28\%   \\
LightFM  & 38\%    & 35\%  & 37\%  & 28\% & 100\%     \\
\bottomrule
\end{tabular}
\\ Source: own study
\end{table}

\subsection{Evaluation of efficiency}

The last step of laboratory evaluation concerned efficiency. For this, we used the following setup: AWS Sagemaker ml.m5.4xlarge instance with 64 GB RAM, 16 vCPUs Intel(R) Xeon(R) Platinum 8259CL CPU @ 2.50GHz running on Amazon Linux AMI 2018.03.

\begin{figure*}[ht]
    \centering
      \resizebox{.9\textwidth}{!}{
    \includegraphics{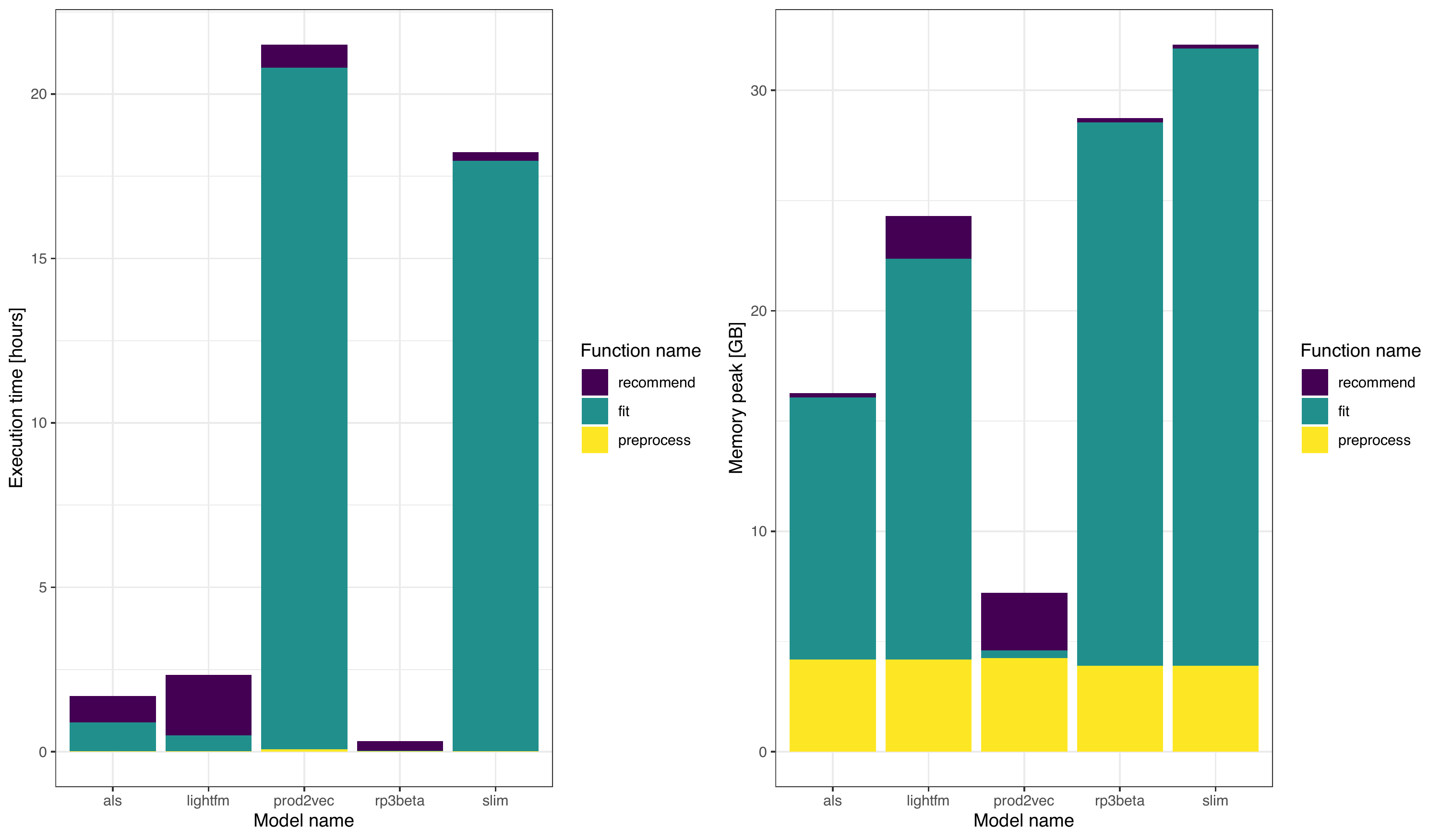}}
    \caption{Execution time and memory utilization. Best viewed in color. Source: own study}
    \label{precision}
  \end{figure*}

In this experiment, all methods were subject to evaluation with respect to three functions: data preprocessing, model fitting, and recommendation of items. 
The execution time was calculated for the total of time needed for the computation of the model (in hours). The memory peak was calculated in GB and concerns maximum memory usage during execution of a given function. This evaluation was performed using the tracemalloc library.\footnote{\url{https://docs.python.org/3/library/tracemalloc.html}, accessed: 2022-08-06} The results of efficiency evaluation are presented in Figure \ref{precision}. We observe that RP3beta, ALS and LightFM outperform SLIM and Prod2Vec in terms of execution time, whereas Prod2Vec requires the least amount of memory.

\subsection{Summary}
The laboratory research enabled us to select models to be tested in an online setting with OLX users.
Firstly, we chose the RP3Beta model for online tests, because it is the best approach in terms of accuracy and execution time. The relatively high memory consumption is still low enough for the model to be tested in production. Investigating the impact of the relatively low diversity of this method may be an interesting line of future research.

The second best model in terms of precision@10 is SLIM. Nevertheless, we did not select this method for online evaluation, due to the low efficiency and the high overlap with the already selected RP3Beta model.
Improving the efficiency of the SLIM implementation might possibly change this decision.

Among other models, the highest precision@10 is achieved by ALS. Additionally, ALS was already implemented and used for job recommendations at OLX; therefore, we decided to include it in the online comparison.

Even though Prod2Vec and LightFM produce the most diverse recommendations, we decided not to test them online due to the significantly worse precision@10, and efficiency in the case of Prod2Vec.

\section{Online evaluation}

To evaluate the effectiveness of the selected methods, we conducted two online A/B tests with users of OLX Jobs. The first step in our experiment served to answer the question: does the introduction of the recommendation method increase the number of users applying for jobs? We focused on users who had recently visited job ads on our platform. In this test, we split the users into two groups: a control group without any recommendations, and a group with recommendations generated using the ALS model.

The test was carried out for 25 days in March 2021. We sent emails with 10 recommended job ads, and when a user had installed an OLX application, a push notification with one recommended job ad was also sent. Then, we observed whether the user applied for any job within the next 48 hours (converted user). The results are reported in Table \ref{involvement}. The statistical significance of the advantage of ALS over the control group was checked using the chi-squared test, and the $p$-value was found to equal 0. This experiment confirmed our hypothesis that recommendations affect activity, as the percentage of converted users after receiving a communication increased significantly, by $(16.83\%-15.98\%)/15.98\% \approx 5.3\%$. The observed rise translates into hundreds of additional users applying for jobs through OLX Jobs daily.

\begin{table}[ht]
\footnotesize
    \centering
\caption{Comparison of activity with or without recommendation.}
\label{involvement}
\begin{tabular}{lll}
\toprule
Variant & Users & \% converted users \\
\midrule
control & 129,308   & 15.98\%            \\
ALS    & 1,170,262  & 16.83\%         \\  
\bottomrule
\end{tabular}
\\ Source: own study
\end{table}

In the second experiment, we tested whether the best model from laboratory testing outperformed ALS.
We split our users into three groups, receiving recommendations from different recommendation systems: ALS, RP3beta, and a mixed variant where  we generated half of the recommendations with ALS and the other half with RP3beta. The second experiment was carried out for 28 days in March and April 2021. The results are presented in Table \ref{tab:message}. The percentage of converted users is calculated here relative to all users; however, changing the recommendation system did not affect users who did not open the message. We decided to filter these out, as their behavior added unnecessary noise to our data. In Table \ref{tab:open_message} we present the impact on users who opened the message. The advantage of RP3beta over ALS is statistically significant ($p$-value $\approx 10^{-6}$). The difference between the variants RP3beta and ALS+RP3beta is not statistically significant (the $p$-value is 0.19).

\begin{table}[ht]
\footnotesize
    \centering
  \caption{Comparison of different recommendation methods.}
  \label{tab:message}
\begin{tabular}{lll}
\toprule
Variant     & Users & \% converted users \\
\midrule
ALS         & 343,892   & 15.25\%            \\
RP3beta     & 345,273   & 15.40\%            \\
ALS+RP3beta & 343,896   & 15.30\%       \\    
\bottomrule
\end{tabular}
\\ Source: own study
\end{table}


\begin{table}[ht]
\footnotesize
    \centering
  \caption{Comparison of different recommendation methods for users who opened the message.}
  \label{tab:open_message}
\begin{tabular}{lll}
\toprule
Variant     & Users & \% converted users \\
\midrule
ALS         & 44,775    & 19.66\%            \\
RP3beta     & 46,097    & 20.94\%            \\
ALS+RP3beta & 45,469    & 20.59\%          \\
\bottomrule
\end{tabular}
\\ Source: own study
\end{table}

Out of users who opened the message, we observe 1.28 pp (percentage points) more converted users from RP3beta than from ALS. Since the message open rate was 13.2\%, this is equivalent to an increase of $1.28\cdot0.132 \approx 0.17$ pp of all targeted users. Based on the results of the first experiment, ALS brings a 0.85 pp increase over the control group. This means that replacing ALS with RP3beta increases our impact on all target users by approximately $0.17/0.85=20\%$.

\subsection{Discussion}
\label{section:discussion}
We have proved that sending job recommendations from the ALS model increases the number of users responding to job ads by more than 5\%. Moreover, RP3Beta significantly outperforms ALS. 

Additionally, we carried out another online A/B test comparing the RP3Beta model with a variant where we replaced half of the RP3Beta recommendations with recommendations generated from the deep learning hybrid recommender system utilized at OLX.\footnote{\url{https://tech.olx.com/item2vec-neural-item-embeddings-to-enhance-recommendations-1fd948a6f293}, accessed: 2022-08-06} We decided not to describe that test in this paper, since this work is focused on collaborative filtering approaches. Additionally, we wanted to keep our results fully reproducible on the published dataset. Unfortunately, we could not publish all of the features utilized by our internal model. We wish merely to note here that we did not observe any improvements over RP3Beta, even though the second model utilizes additional knowledge about job ads. Hence, we confirmed that a properly determined and tuned simple model may outperform much more advanced deep solutions. This is consistent with the results from \cite{Dacrema2019} and \cite{reenvisioningNCF}, who compared several non-neural approaches, including RP3Beta, with more advanced neural models.

The efficiency of the RP3Beta model enables OLX Jobs to train it from scratch multiple times per day with very low costs (on a CPU). It is currently the state-of-the-art collaborative filtering model at OLX Jobs. We are working on improving the model without decreasing the efficiency drastically, which is a very important restriction in real-world applications. We observe a tendency for other companies also to utilize simpler models; for example, Netflix decided not to implement the winning algorithm of the Netflix Prize competition \citep{bennett2007netflix}, because of the too great engineering effort.\footnote{\url{https://towardsdatascience.com/deep-dive-into-netflixs-recommender-system-341806ae3b48}, accessed: 2022-08-06}

In this work, we have filled the research gap consisting in a lack of comparison of classical collaborative filtering approaches in the jobs domain, by providing comprehensive laboratory research on real-world dataset and online evaluation with millions of users. Additionally, we have made the source code for offline evaluation and the dataset publicly available. We encourage other researchers to develop new recommendation methods, which we hope will help users find their ideal job more quickly.

\section{Summary}

This paper has addressed the topic of job recommendations in online classifieds, using the example of OLX Jobs. We performed an extensive evaluation of approaches representing different families of recommendation methods.

During the laboratory research, we demonstrated that RP3beta, SLIM, and ALS perform significantly better than LightFM and Prod2Vec. We also found that Prod2Vec offers the greatest diversity, which may be important when working on the novelty of recommendations. Moreover, we checked how similar are the recommendations generated by different models.

We also performed an evaluation by sending millions of messages to online users. A/B tests were carried out using ALS and RP3beta.
The results of these tests demonstrated that sending job recommendations generated by these models increases the number of users contacting advertisers from OLX Jobs. It needs to be noted that with RP3beta available in the production environment, we improve our effectiveness by 20\% (when considering reactions of users).

When evaluating the methods from the classifieds perspective, we need to note that a model recommending the most popular (as yet unseen) items is not suitable for job recommendations. 
A recommendation model based on random recommendations was also added for comparison to demonstrate the advantage of the researched methods.

Another important result of this research is the dataset, which will make it possible to carry out more research on recommendations for the needs of classifieds.

Future work related to our research will include different interaction types to differentiate user preferences towards an item (e.g., viewing or response to an ad) and the frequency of visiting the same ad multiple times. Such research is made possible by the dataset which we have introduced in this work.

\newpage

\bibliographystyle{elsarticle-harv}\biboptions{authoryear}
\bibliography{bib}

\end{document}